\documentclass{PoS}
\usepackage{amsmath}

\title{Effective theories and resonances in strongly-coupled electroweak symmetry breaking scenarios}

\ShortTitle{Effective theories and resonances in strongly-coupled EWSB scenarios}

\author{\speaker{Ignasi Rosell}\thanks{We wish to thank the organizers for the pleasant conference. This work has been supported in part by the Spanish Government and ERDF funds from the European Commission (FPA2016-75654-C2-1-P, FPA2017-84445-P); by the Spanish Centro de Excelencia Severo Ochoa Program (SEV-2014-0398); by the Generalitat Valenciana (PROMETEO/2017/053); by the Universidad Cardenal Herrera-CEU (INDI17/11 and INDI18/11); and by the STSM Grant from COST Action CA16108. C.K. acknowledges the support of the Alexander von Humboldt Foundation. This manuscript has been authored by Fermi Research Alliance, LLC under Contract No. DE-AC02-07CH11359 with the U.S. Department of Energy, Office of Science, Office of High Energy Physics. Preprint numbers: IFIC/19-38, FTUV/19-1004, FERMILAB-CONF-19-479-T}\\
        Departamento de Matem\'aticas, F\'\i sica y Ciencias Tecnol\' ogicas, Universidad Cardenal Herrera-CEU, CEU Universities, 46115 Alfara del Patriarca, Val\`encia, Spain \\
        E-mail: \email{rosell@uchceu.es}}
        
\author{Claudius Krause \\
Theoretical Physics Department, Fermi National Accelerator Laboratory, Batavia, IL, 60510, USA \\   E-mail: \email{ckrause@fnal.gov}}        
        
\author{Antonio Pich  \\
IFIC, Universitat de Val\`encia -- CSIC, Apt. Correus 22085, 46071 Val\`encia, Spain\\   E-mail: \email{pich@ific.uv.es}}
    
    \author{Juan Jos\'e Sanz-Cillero \\ Departamento de F\'\i sica Te\'orica,  Universidad Complutense de Madrid, E-28040 Madrid, Spain \\ E-mail: \email{jjsanzcillero@ucm.es}}

\abstract{Due to the mass gap between the Standard Model and possible New Physics states, electroweak effective approaches are appropriate. Although a linear realization of the electroweak symmetry breaking with the Higgs forming a doublet together with the Goldstone bosons of the EWSB is a first possibility (SMEFT), we adopt the more general non-linear realization, where the Higgs is a singlet with independent couplings (EWET, HEFT or EWChL). %Note that the EWET includes the SMEFT as a particular case. 
We present the effective Lagrangian at low energies (the EWET, with only the SM fields) and at high energies (the resonance theory, with also a set of resonances). Taking into account the high scale of these resonances, their experimental searches seem to be more accessible by considering their imprints at low-energies, {\it i.e.}, their imprints in the Low Energy Constants (LECs) of the EWET at energies lower than the resonance masses. We give some examples of these phenomenological connections.}

\FullConference{
European Physical Society Conference on High Energy Physics - EPS-HEP2019 -\\
			10-17 July, 2019\\
			Ghent, Belgium}

\begin{document}

\section{Introduction}

The LHC continues to confirm the success of the Standard Model (SM): the Higgs-like particle couples following the SM predictions and searches for New Physics (NP) have given negative results so far. The consequent mass gap between the electroweak (EW) and possible NP scales allows us to use effective field theories (EFTs) at current energies and, moreover, they give a systematic way to analyze the imprints of possible new particles at low energies: the study of the low-energy constants (LECs) of the next-to-leading (NLO) effective Lagrangian.

If the LECs are considered free parameters, EFTs are model-independent but for some well-motivated assumptions: particle content, symmetries and power counting. The particle content is clear in this case, the SM fields, but there are two different ways of introducing the Higgs, and this choice has consequences in the symmetries and in the power counting to be used. One can consider the Higgs forming a doublet with the three Goldstone bosons of the electroweak symmetry breaking (EWSB) or without assuming any specific relation between the Higgs and the Goldstone bosons. The first one, usually called SM effective field theory (SMEFT), corresponds to the linear realization of the EWSB and it is an expansion in canonical dimensions, whereas the second one, usually called EW effective theory (EWET), EW chiral Lagrangian (EWChL) or Higgs effective field theory (HEFT), is a more general (non-linear) realization of the EWSB, it is an expansion in generalized momenta and it includes the SMEFT as a particular case. We follow the second option.

At high energies, where the resonances are supposed to live, these NP states can be incorporated by using a phenomenological Lagrangian respecting the EWSB pattern of the SM: the resonance theory. We want to stress that we are considering two effective Lagrangians for two different regimes: the EWET at low energies and the resonance theory at high energies. Once there is a big energy gap between the SM states and possible resonances, it makes sense to integrate out the resonances, recovering the EWET Lagrangian with the LECs given in terms of resonance parameters. The phenomenology we present here is based on these ideas.

\section{The Lagrangians}

\subsection{Low energies: the electroweak effective theory}

The construction of the EWET Lagrangian can be found in Ref.~\cite{EWET} and its main ingredients are the following ones: the particle content of the SM (with the Higgs as as a scalar singlet, as it has been pointed out previously); the EWSB pattern, $\mathcal{G}\equiv SU(2)_L\otimes SU(2)_R \rightarrow \mathcal{H}\equiv SU(2)_{L+R}$; and an expansion in powers of generalized momenta: $\mathcal{L}_{\mathrm{EWET}} = \mathcal{L}_{\mathrm{EWET}}^{(2)}+\mathcal{L}_{\mathrm{EWET}}^{(4)} +\dots$. The LO Lagrangian, $\mathcal{L}_{\mathrm{EWET}}^{(2)}$, is basically the SM Lagrangian with a general scalar singlet Higgs, whereas the NLO Lagrangian can be organized considering the parity ($P$) of the operators and the number of fermion bilinears~\cite{EWET}:
\begin{equation}
  \label{eq:L-NLO}
\mathcal{L}_{\mathrm{EWET}}^{(4)}  =
\sum_{i=1}^{12} \mathcal{F}_i\; \mathcal{O}_i  + \sum_{i=1}^{3}\widetilde{\mathcal{F}}_i\; \widetilde{\mathcal{O}}_i  
 +  \sum_{i=1}^{  8  } \mathcal{F}_i^{\psi^2}\; \mathcal{O}_i^{\psi^2}  + \sum_{i=1}^{  3   } \widetilde{\mathcal{F}}_i^{\psi^2}\; \widetilde{\mathcal{O}}_i^{\psi^2}
 + \sum_{i=1}^{10}\mathcal{F}_i^{\psi^4}\; \mathcal{O}_i^{\psi^4}  + \sum_{i=1}^{2}\widetilde{\mathcal{F}}_i^{\psi^4}\; \widetilde{\mathcal{O}}_i^{\psi^4}  ,
\end{equation}
where operators without (with) tilde are $P$-even ($P$-odd) and where we split the Lagrangian into bosonic, two-fermion and four-fermion operators, respectively.

\subsection{High energies: the resonance theory} \label{RET_section}

We consider now the particle content of the SM plus bosonic fields with $J^P=0^\pm$ and $J^P=1^\pm$ (in electroweak triplets or singlets and in QCD octets or singlets) and fermionic states with $J=\frac{1}{2}$ (in electroweak doublets and in QCD triplets or singlets)~\cite{EWET}. The power counting of the EWET is not directly applicable here, but one can follow a consistent organization taking into account the integration of the resonances (in general $R\propto1/M_R^2$ and $\Psi\propto1/M_\Psi$ for the classical solution of the bosonic and fermionic resonances) and the additional suppression in case of fermionic fields due to their weak coupling: only operators containing up to one resonance and constructed with chiral operators of $\mathcal{O}(p^2)$ or lower are needed. 

Considering the mass gap between possible heavy resonances and the SM particles, the resonances can be integrated out and the EWET LECs are given in terms of resonance parameters, {\it i.e.}, $\mathcal{F}_i,\,\widetilde{\mathcal{F}}_i,\,\mathcal{F}_i^{\psi^2}, \widetilde{\mathcal{F}}_i^{\psi^2},\,\mathcal{F}_i^{\psi^4}$ and $\widetilde{\mathcal{F}}_i^{\psi^4}$ in (\ref{eq:L-NLO}) in terms of resonance couplings and masses~\cite{EWET}.

\section{Phenomenology}

\subsection{By using the $S$ and $T$ parameters} \label{ST_section}

The existence of massive resonances coupled to the gauge bosons would modify the $Z$ and $W^\pm$ self energies and these deviations with respect to the SM are properly characterized by the parameters $S$ and $T$~\cite{Peskin:1990zt}. In Ref.~\cite{S_T} a determination of $S$ and $T$ at NLO with a simplified version of the resonance theory presented above was given:
% we determined $S$ and $T$ at NLO with a simplified version of the resonance theory presented above: %only colorless vector and axial-vector resonances were present in the resonance Lagrangian (neither scalar nor pseudoscalar nor fermionic resonances); 
%we assumed 
the absence of $P$-odd operators was assumed and only the one-loop contributions from the lightest two-particle bosonic channels (two Goldstones or one Goldstone plus the Higgs) were considered. In order to reduce the number of resonance parameters %and, consequently, to get phenomenological results
high-energy constraints were considered: the assumption of well-behaved form factors and the Weinberg Sum Rules (WSRs)~\cite{WSR} for the $W^3B$ correlator. A generalization of this analysis using also $P$-odd operators and considering the two-particle fermionic channels too is in preparation~\cite{future}.

\begin{figure}[t]
\centering
\includegraphics[scale=0.55]{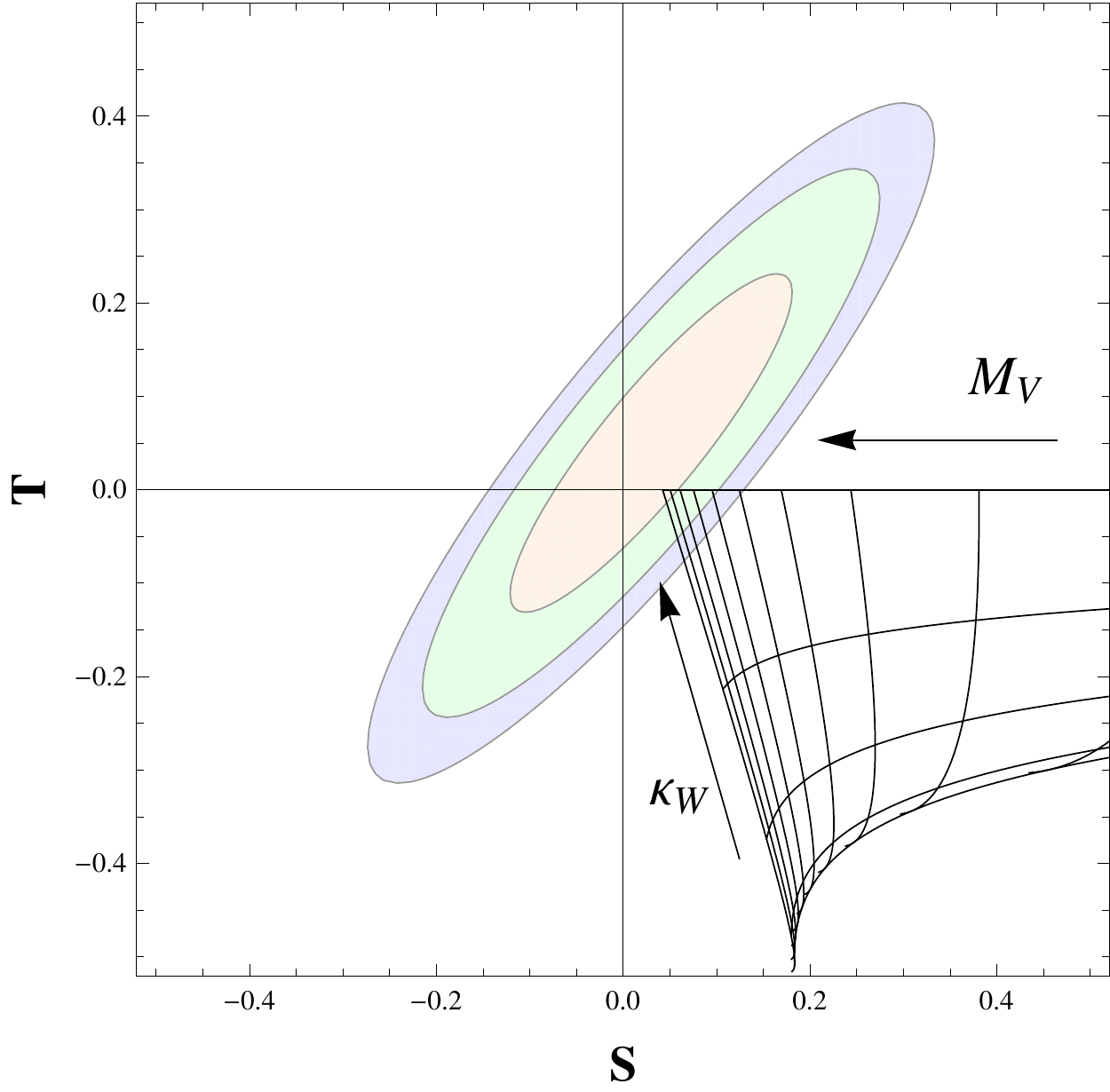} \quad \includegraphics[scale=0.60]{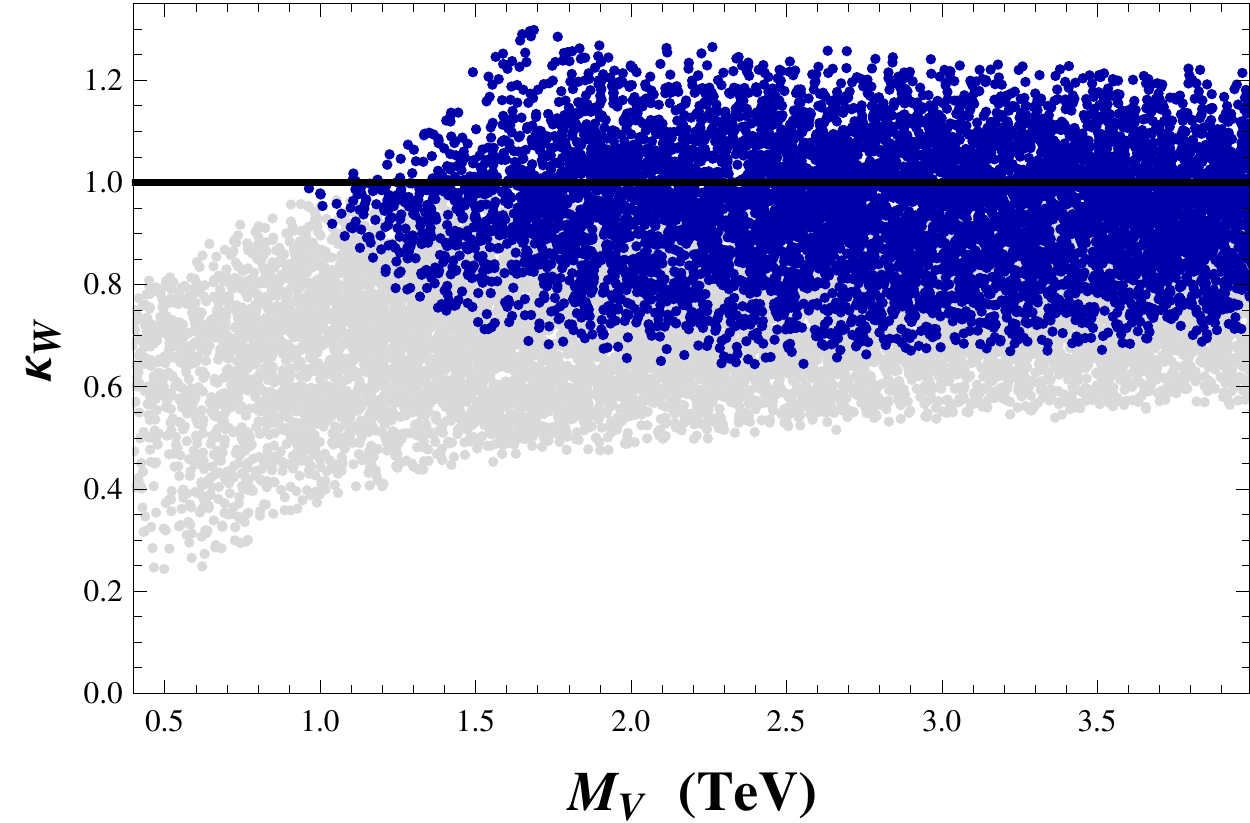}
\caption{{\bf NLO determinations of $S$ and $T$, imposing the two WSRs (left)}.
The approximately vertical curves correspond to constant values of $M_V$, from $1.5$ to $6.0$~TeV at intervals of $0.5$~TeV. The approximately horizontal curves have constant values of $\kappa_W$: $0.00, \, 0.25, 0.50, 0.75, 1.00$. The ellipses give the experimentally allowed regions at 68\%, 95\% and 99\% CL. {\bf Scatter plot for the 68\% CL region, in the case when only the first WSR is assumed (right)}. The dark blue and light gray regions correspond, respectively,  to $0.2<M_V/M_A<1$ and $0.02<M_V/M_A<0.2$. The experimental bounds are extracted from Ref.~\cite{Baak:2012kk}.} \label{fig-1}       
\end{figure}

With all these ingredients results up to NLO in terms of only a few parameters were shown in Ref.~\cite{S_T}: assuming the first and the second WSRs, $S$ and $T$ were given in terms of $M_V$ and the hWW coupling $\kappa_W$ (or, equivalently $M_A$, since $\kappa_W=M_V^2/M_A^2$ in this case) and, assuming only the first WSR, $T$ and a lower bound of $S$ in terms of $M_V$, $M_A$ and $\kappa_W$. The results are shown in Figure~\ref{fig-1}~\cite{S_T}: the masses of vector and vector-axial resonances were pushed to the TeV range, their mass splitting is supposed to be small and %the Higgs-like boson requires a $WW$ coupling close to the SM one ($\kappa_W =1$).
$\kappa_W$ is required to be close to the SM value ($\kappa_W =1$).

\subsection{By estimating the bosonic LECs}

\begin{figure*}[t]
\begin{center}
%\begin{center}
\begin{minipage}[c]{5.5cm}
\includegraphics[width=5.5cm]{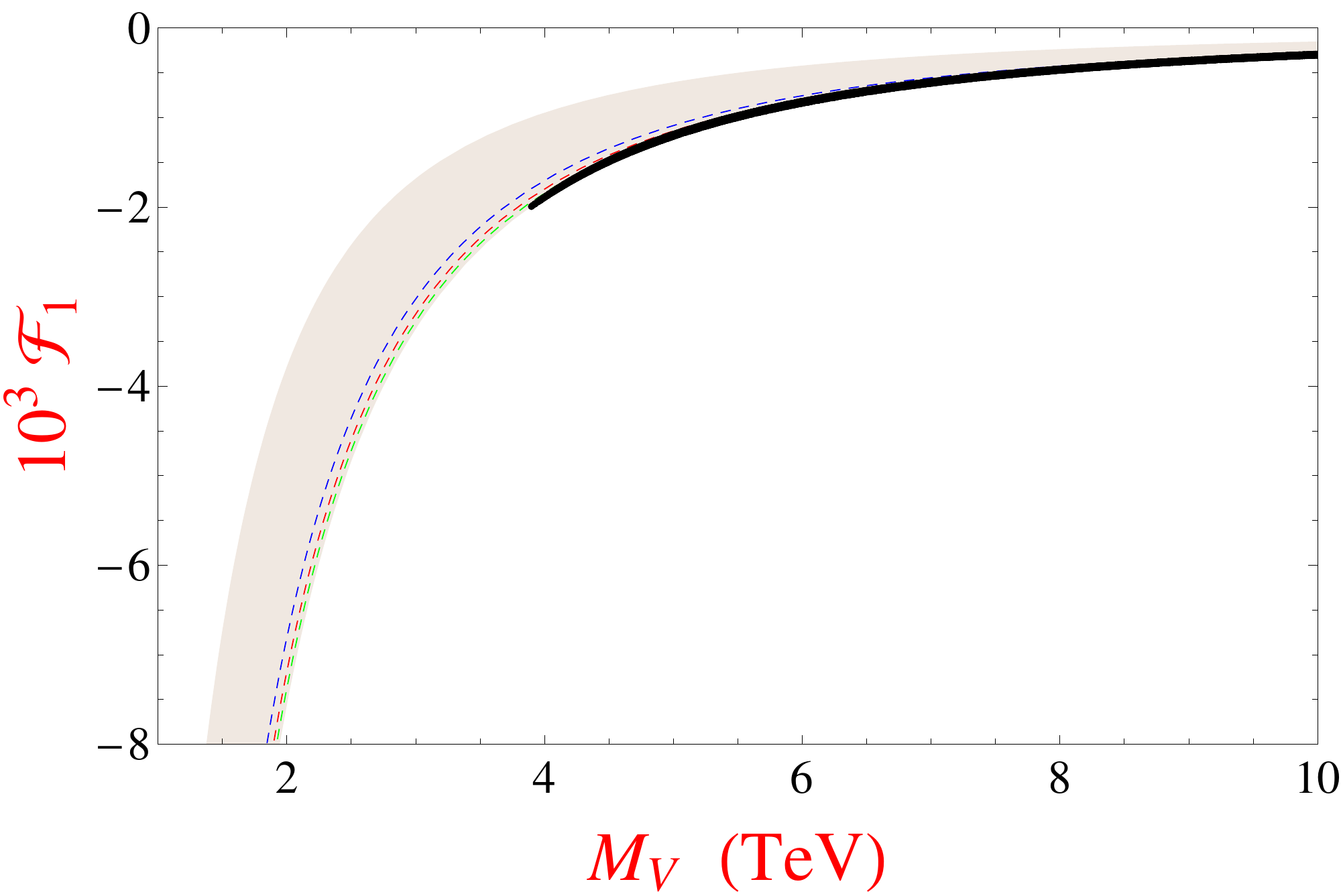}
\end{minipage}
\hskip .5cm
\begin{minipage}[c]{5.5cm}
\includegraphics[width=5.5cm]{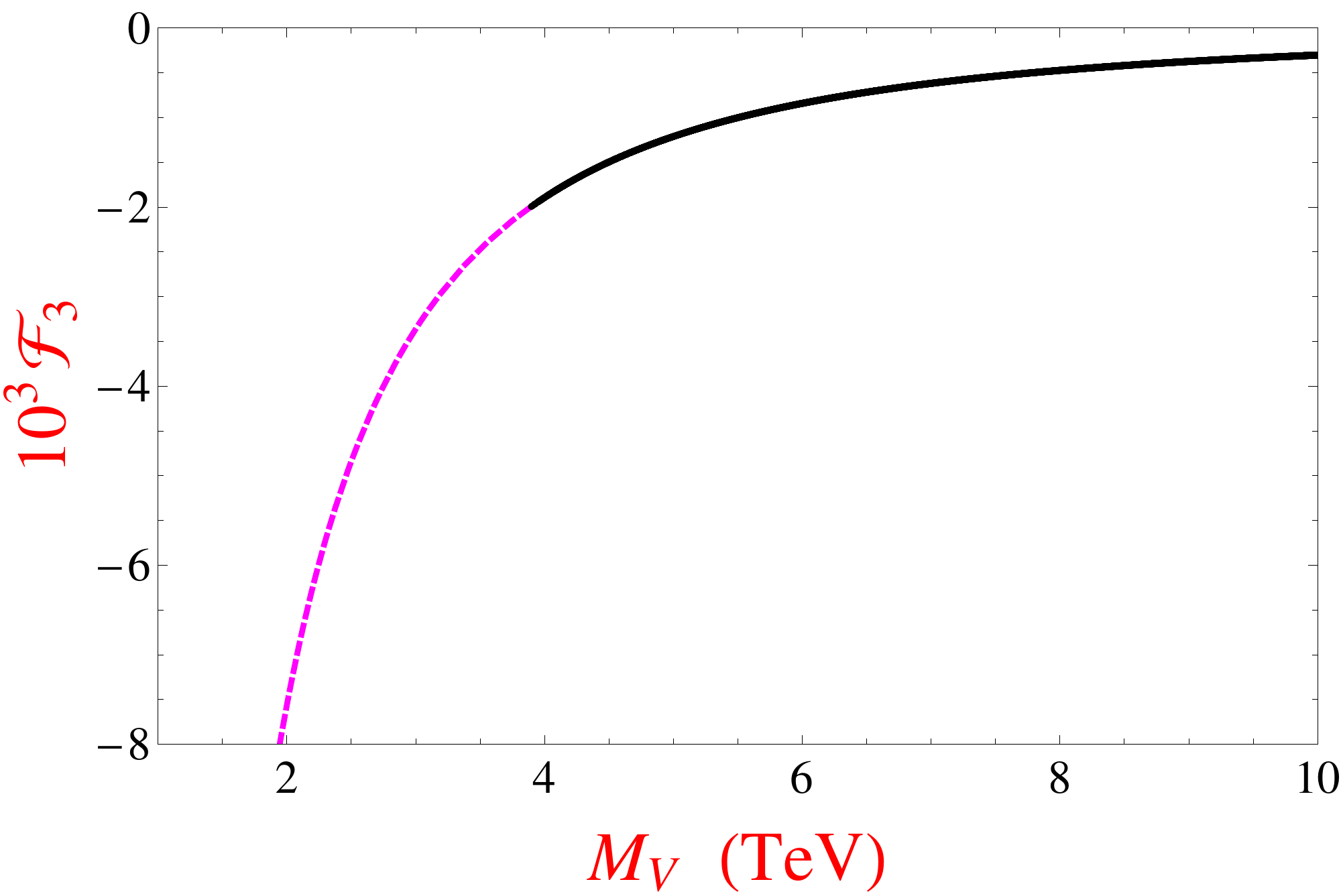}
\end{minipage}
\\[8pt]
\begin{minipage}[c]{5.5cm}
\includegraphics[width=5.5cm]{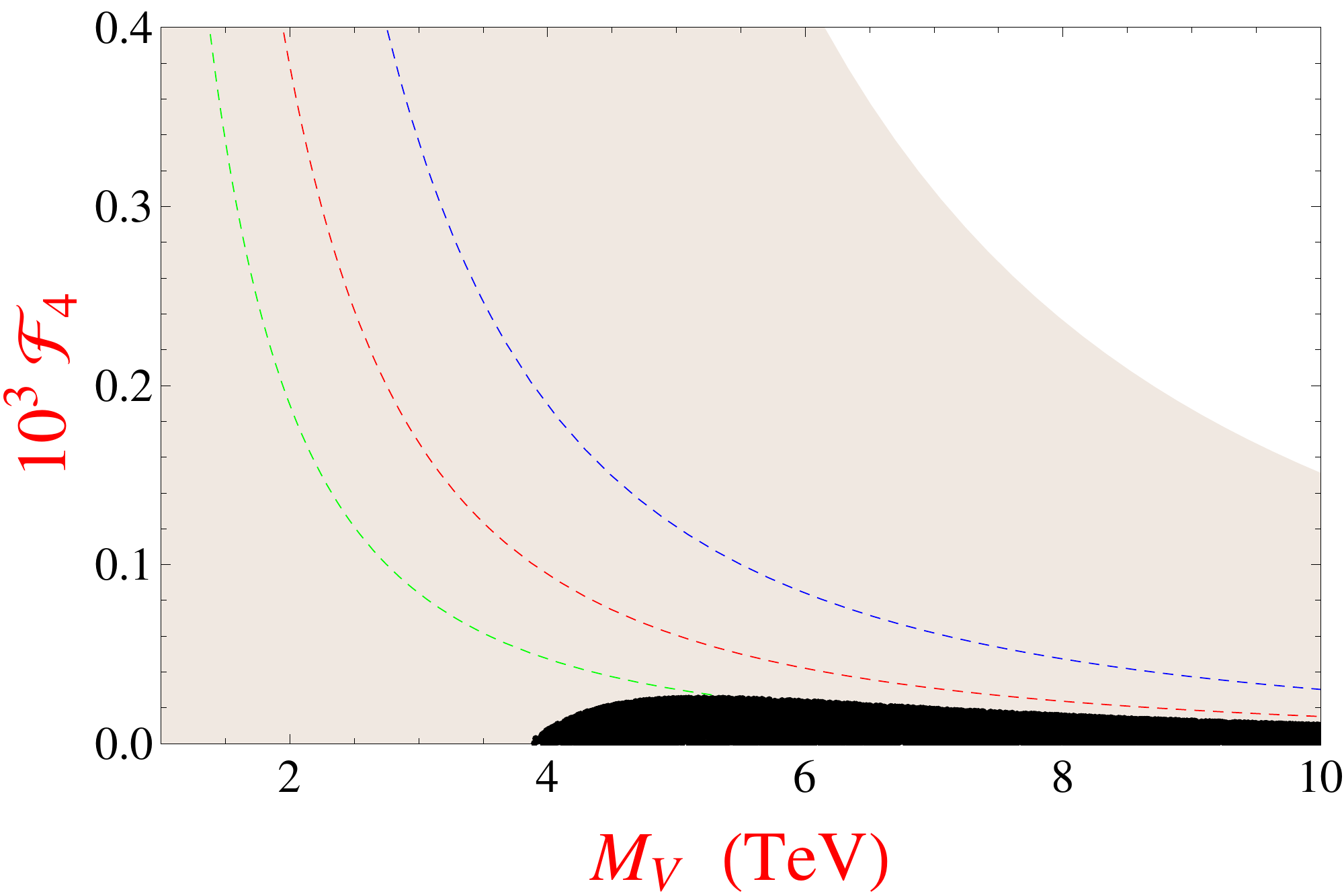}
\end{minipage}
\hskip .2cm
\begin{minipage}[c]{5.5cm}
\includegraphics[width=5.5cm]{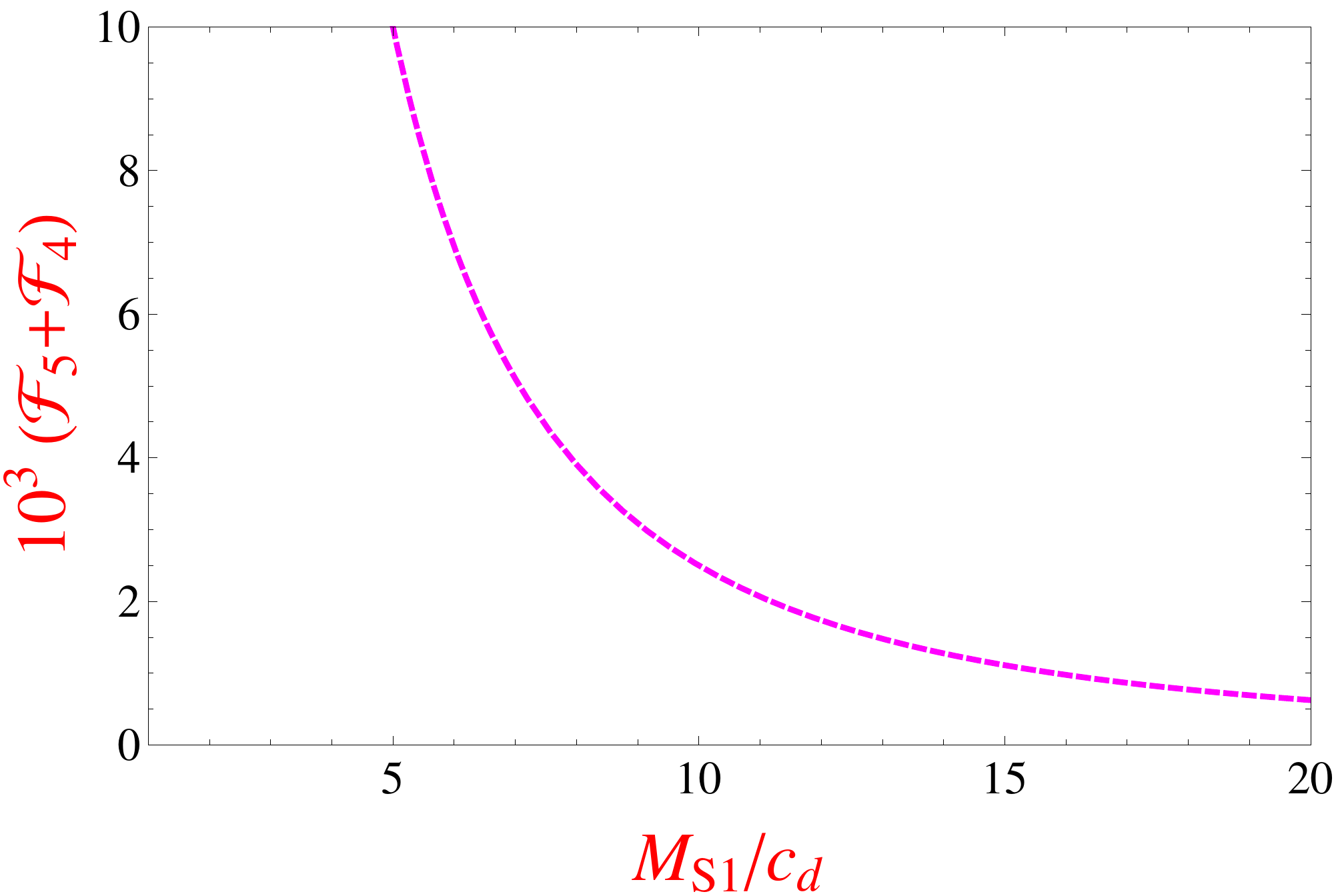}
\end{minipage}
\caption{Predicted LECs as function of resonance masses. The light-shaded regions cover all possible values for $M_A>M_V$, while the blue, red and green lines correspond to $\kappa_W= M_V^2/M_A^2 = 0.8,\, 0.9$ and 0.95, respectively. $\mathcal{F}_3$ does not depend on $M_A$, whereas $\mathcal{F}_4+\mathcal{F}_5$ only depends on $M_{S_1}/c_d$. The oblique $S$ and $T$ constraints restrict the allowed ranges (95\% C.L.) to the dark areas.}
\label{fig-2}   
\end{center}
\end{figure*}

As a first step in the integration of the resonances explained in Section~\ref{RET_section}~\cite{EWET}, the $P$-even operators of the resonance theory were considered in Ref.~\cite{Pich:2015kwa} to determine most of the purely bosonic LECs of the EWET, $\mathcal{F}_{1-9}$ in (\ref{eq:L-NLO}), in terms of %a few
resonance parameters:% The results are shown in Table~\ref{table1} and, 
\begin{align}
\mathcal{F}_1&=-\frac{v^2}{4}\,\left(\frac{1}{M_V^2}+\frac{1}{M_A^2}\right)\,, &  
\mathcal{F}_2&=-\frac{v^2 (M_V^4+M_A^4)}{8 M_V^2M_A^2 (M_A^2-M_V^2)}\,, & 
\mathcal{F}_3&=-\frac{v^2}{2M_V^2}\,, \nonumber \\
\mathcal{F}_4&= \frac{(M_A^2-M_V^2) v^2}{4 M_V^2 M_A^2},&
\mathcal{F}_5&= \frac{c_{d}^2}{4M_{S_1}^2} -\frac{(M_A^2-M_V^2) v^2}{4 M_V^2 M_A^2},&
\mathcal{F}_6&= -\frac{M_V^2 (M_A^2-M_V^2) v^2}{M_A^6},  \nonumber \\
\mathcal{F}_7&= \frac{d_P^2}{2 M_P^2}+\frac{M_V^2 (M_A^2-M_V^2) v^2}{M_A^6},&
\mathcal{F}_8&= 0,&
\mathcal{F}_9&= -\frac{M_V^2 v^2}{M_A^4}. 
\end{align}
Note that, and as in the previous subsection, the use of short-distance constraints was fundamental to be able to give the LECs in terms of only a few resonance parameters (well-behaved form factors and WSRs again). In Figure~\ref{fig-2}~\cite{Pich:2015kwa} we show the numerical values of the most interesting $\mathcal{F}_i$ (from a phenomenological point of view). The light-shaded regions indicate all a priori possible values for $M_A > M_V$. The dashed blue, red and green lines correspond to $\kappa_W=M_V^2/M_A^2 = 0.8,\, 0.9$ and 0.95, respectively. Note that $\mathcal{F}_3$ does not depend on $M_A$, whereas $\mathcal{F}_4+\mathcal{F}_5$ only depends on $M_{S_1}/c_d$. 

\begin{figure}[t]
\begin{center}
\centering
\includegraphics[scale=0.29]{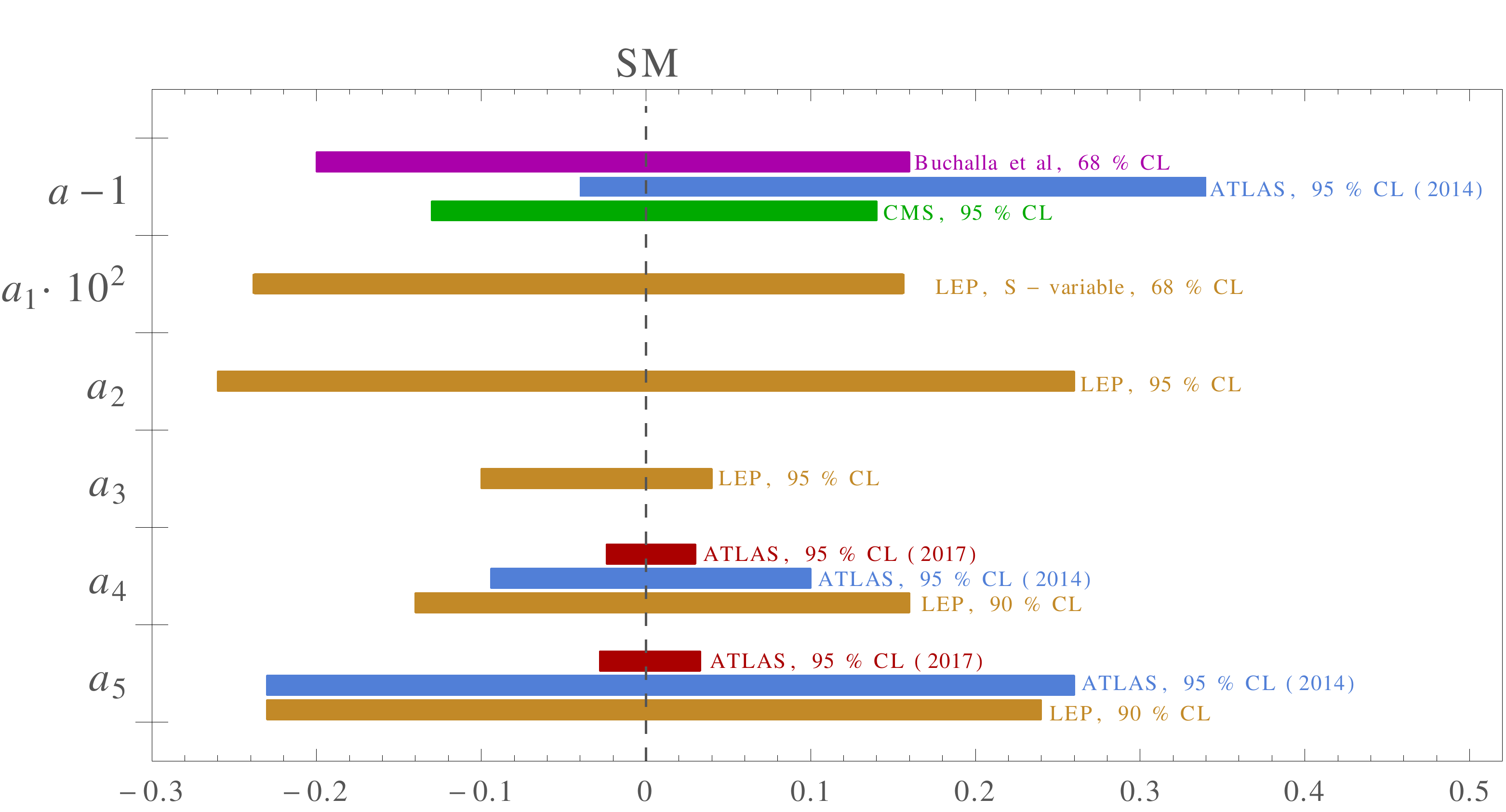} % I prefer scale=0.32
\caption{Experimental constraints on bosonic LECs of the EWET Lagrangian. Figure from Ref.~\cite{Delgado:2017cls}. The constraints of $a_1$ have been updated. The constraints are extracted from Ref.~\cite{phen}. } \label{fig-3}       
\end{center}
\end{figure}

%A figure with experimental bounds on these LECs appears in Ref.~\cite{Delgado:2017cls}, Figure~\ref{fig-3}.%, where the constraints are extracted from Refs.~\cite{phen}. 
A summary of experimental bounds on these LECs was given in Ref.~\cite{Delgado:2017cls}, Figure~\ref{fig-3} in these proceedings.
The original notation of Ref.~\cite{Longhitano:1980iz} is followed in the figure, whose translation to our notation is given by $a=\kappa_W$, $a_1=\mathcal{F}_1$, $a_2-a_3=\mathcal{F}_3$, $a_4=\mathcal{F}_4$ and $a_5=\mathcal{F}_5$. 

As in the previous subsection, and considering the theoretical predictions shown in Figure~\ref{fig-2}~\cite{Pich:2015kwa} and the experimental constraints of Figure~\ref{fig-3}~\cite{Delgado:2017cls}, lower bounds in the resonance masses are of the order of the TeV scale. 

%%%%%%%%%%%%%%%%%%%%%%%%%%%%%%%%%%%%%%%%%%
%%%%%%%%%%%%%%%%%%%%%%%%%%%%%%%%%%%%%%%%%%

\subsection{By searching for four-fermion operators}

Searches for four-fermion operators as the ones of (\ref{eq:L-NLO}) (whose LECs $\mathcal{F}_i^{\psi^4}$ and $\widetilde{\mathcal{F}}_i^{\psi^4}$, accordingly, are not present in the SM) can be found in the literature and can be used to find bounds for NP scales. Standard dijet and dilepton studies at LHC and LEP have looked for four-fermion operators containing light quarks and/or leptons~\cite{LHC_LEP}. In these studies, one expresses generally the LECs in terms of a NP scale $\Lambda$ defined through $|\mathcal{F}^{\psi^4}_j| = 2\pi/\Lambda^2$. Stringent (95\% CL) lower limits on this scale read~\cite{EWET}:%
\begin{enumerate}
\item From dilepton production~\cite{LHC_LEP}: $\Lambda\geq 24.6$~TeV from LEP, $\Lambda\geq 19.0$~TeV from CMS and $\Lambda\geq 26.3$~TeV from ATLAS.
\item From dijet production~\cite{LHC_LEP}: $\Lambda\geq 16.2$~TeV from LEP, $\Lambda\geq 18.6$~TeV from CMS and $\Lambda\geq 21.8$~TeV from ATLAS.
\end{enumerate}
The previous bounds refer to four-fermion operators with light quarks (first and second generation). There are studies including also top and bottom quarks~\cite{top_bottom}:
\begin{enumerate}
\item From low-energy analysis~\cite{top_bottom}: $\Lambda\geq 3.3$~TeV from semileptonic B decays and $\Lambda\geq 14.5$~TeV from $B_s-\overline{B}_s$ mixing.
\item From high-energy collider analysis~\cite{top_bottom}: $\Lambda\geq 2.3$~TeV from $t$ and $t\bar{t}$ production at LHC and Tevatron, $\Lambda\geq 1.5$~TeV from multi-top production at LHC and Tevatron and $\Lambda\geq 4.7$~TeV from dilepton production at LHC.
\end{enumerate}

Note that, as in the previous subsections, NP scales are pushed to TeV scales~\cite{EWET}. Consequently, precise phenomenology at low energies by using EFTs (SMEFT or the more general EWET) continues to be a good way to search for New Physics~\cite{future}.

\end{document}